# Deep levels analysis in wavelength extended InGaAsBi photodetector

Jian Huang,[1] Baile Chen,[1,a)] Zhuo Deng,[1] Yi Gu,[2,3] Yingjie Ma,[2,3] Jian Zhang,[2] Xiren Chen, [4]and Jun Shao, [4]

[1] *Optoelectronic Device Laboratory, School of Information Science and Technology, ShanghaiTech University, Shanghai 201210, China*

[2] *State Key Laboratory of Functional Materials for Informatics, Shanghai Institute of Microsystem and Information Technology, Chinese Academy of Sciences, Shanghai 200050, China*

[3] *State Key Laboratory of Transducer Technology, Shanghai Institute of Technical Physics, Chinese Academy of Sciences, Shanghai 200083, China*

[4] *National Laboratory for Infrared Physics, Shanghai Institute of Technical Physics, Chinese Academy of Sciences, 200083 Shanghai, China*

## ABSTRACT

InP based dilute Bismide InGaAsBi material is emerging as a promising candidate for extending short wavelength infrared detection. One critical factor to limit the performance of these InGaAsBi photodiodes is dark current caused by defects within the material. In this work, low frequency noise spectroscopy (LFNS) and temperature varied photoluminescence was used to characterize the defect levels in the devices. Three deep levels located at $E_c$ -0.33 eV, $E_v$ +0.14 eV, and $E_c$ -0.51 eV were identified from the LFNS spectra, which are consistent with emission peak energy found by photoluminescence spectra of InGaAsBi.

## INTRODUCTION

InP based InGaAs material system has been widely used in high performance photodetectors (PDs) over past several decades. In particular, the lattice-matched $In_{0.53}Ga_{0.47}As$ PDs on InP has been demonstrated with low dark current, high efficiency and high speed in 1.1µm–1.7µm wavelength range [1]. For the applications in areas such as chemical sensing, gas monitoring and infrared imaging, it would be more desirable to extend the detection wavelength beyond 1.7µm. Longer detection wavelength can be achieved by using Indium-rich InGaAs materials [2-4] or InGaAs/GaAsSb type-II multiple quantum wells (MQW) structures [5-13]. For example, Hamamatsu [14] sells an uncooled In-rich InGaAs PD with cutoff wavelength of 2.6 µm, peak responsivity of 1.3 A/W. Sidhu et.al. [7] initially demonstrated an InP based InGaAs/GaAsSb type-II MWQ PIN photodiodes with cutoff wavelength out to 2.5 µm with a peak detectivity of 3.8 x $10^9$ $cmHz^{0.5}/W$ at room temperature.

1. Corresponding author: chenbl@shanghaitech.edu.cn.

Recently, Bismuth (Bi) containing III-V semiconductor alloys have attracted much attention due to the strong band gap dependence on Bi [15-19]. Several groups have reported that many III-V materials such as InGaAs or GaAs exhibit a strong reduction in bandgap with a small Bi fraction, which is attributed to the anticrossing interaction between the valence band and Bi resonant states [20-22]. Therefore, InGaAsBi is also a promising absorber which can extend wavelength detection beyond 1.7μm on InP [23]. Despite the promising attributes, growth of high quality dilute Bismide InGaAsBi on InP is still challenging and the dark current of InGaAsBi PD is high. The incorporation of Bi in III-V brings deep localizations in the bandgap due to states of isolated Bi atoms resonant within the valence band [24, 25], which act as deep level traps and limit the performance of device. Therefore, it is crucial to have an in-depth characterization and understanding of the deep level traps in InGaAsBi materials.

In this paper, we studied the defect property of the InGaAsBi materials on InP substrate based on temperature varied dark current, optical response, photoluminescence (PL) and low frequency noise spectroscopy (LFNS) measurement. The devices show anomalous red-shift behavior as temperature changes from 77K to 300K, which is discussed in detailed along with the photoluminescence (PL) and low frequency noise spectroscopy analysis. Moreover, three different deep levels found from LFNS with activation energy of 0.33 eV, 0.14 eV and 0.51 eV, respectively, are also consistent to the temperature-independent peak behaviors observed in PL spectra.

## GROWTH AND FABRICATION

The sample was grown on InP substrate with a p-i-n structure by using molecular beam system (MBE). Fig. 1 shows the schematic of the InGaAsBi PD, the growth started with a 1μm thick n-type InP buffer layer, which was heavily doped with Si to $3\times10^{18}$ cm$^{-3}$. After that, Si-doped InGaAsBi absorption layer with doping density of $3\times10^{16}$ cm$^{-3}$ and thickness of 1.5μm was grown, which is followed by a 600 nm thick InP layer heavily doped with Be ($5\times10^{18}$ cm$^{-3}$). Finally, the sample was capped with a 100 nm thick p-type $In_{0.53}Ga_{0.47}As$ layer heavily doped with Be ($5\times10^{18}$ cm$^{-3}$).

The Bi composition in InGaAsBi was measured by high resolution X-ray diffractometer (HRXRD) (004) ω/2θ curves as shown in Fig. 2. The narrow and highest peak comes from InP substrate, and other two weak peaks correspond to InGaAs contact and InGaAsBi absorption layer. The lattice mismatch between InGaAsBi absorption layer and InP substrate is 0.09%, Here the Indium composition is reduced to 50%, and the Bi composition in InGaAsBi is about 3%.

After the material growth, the sample was fabricated into different sizes diameter mesa-shaped PDs. Standard photolithography and wet chemical etching were used to define the mesa. Ohmic contact electrodes were deposited by using electron evaporation and lift-off techniques, $Si_3N_4$ was used to passivate the etched surface.

## ELECTRICAL AND OPTICAL ANALYSE

The dark current-voltage relation of a 300μm diameter PD was measured from 77K-300K in a variable temperature probe station, as shown in Fig. 3 (a). Under the bias voltage of -0.1V at 300K, the dark current is $2.2\times10^{-5}$ A, which corresponding to a dark current density of $3.1\times10^{-2}$ A/cm$^2$. Fig. 3(b) shows the Arrhenius plot of the temperature dependent dark current, the activation energy ($E_a$) is about 0.29 eV at high temperature region (220K-300K), and at low temperature region(77K-160K) the $E_a$ decreases to 0.14 eV. Compare with effective bandgap (~0. 47eV) at 300K of the InGaAsBi obtained from spectral response shown in later section, it suggests that the deep levels related generation-recombination (G-R) current contributes significant to dark current at high temperature region. At low temperature region $E_a$ is lower than the $E_g/3$, suggesting that the dark current could be dominated by tunneling current . However, the activation energy extracted in the Arrhenius plot generally represents the average energy of multiple traps rather than the signature of one single trap. Therefore, it would be necessary to have a more detailed deep level characterizations on the origin of the dark current in these devices.

After the electrical characterization, the device was wire-bonded and loaded into a low temperature cryostat for the optical response measurement. The photo-response of the device was carried out with frontside illumination and without anti-reflection (AR) coating. The photocurrent was amplified and analyzed by a low noise current preamplifier and a fast Fourier transform (FFT) network spectrum analyzer. A ThermoFisher Scientific Nicolet Fourier transform infrared spectrometer (FTIR) was used to measure the relative responsivity of the device, a standard black body source at 700 ℃ was used to calibrate the relative response. Fig. 4 shows the normalized responsivity of the device under zero bias at different temperatures, all the data has been normalized to the max value of the responsivity under 77K. Compare with the InGaAs/InP PD, the cutoff wavelength of the InGaAsBi/InP PD has extended to about 2.25μm at 77K. With the temperature increasing to 300K, the detection wavelength extends to 2.66μm. It is interesting to note that the cutoff wavelength (~2.25μm) of the device is almost temperature-independent under 220K, and no obvious red shift is indicated in Fig. 4. However, as the temperature increases above 220K, the red shift of the cutoff wavelength becomes visible. This abnormal red shift phenomenon as temperature increases can be attribute to the Bi-related temperature-insensitive edge states near the valence band. As the temperature is below 220K, this Bi-related edge state is above valence band, and cutoff wavelength of the device is mainly determined by the energy difference between the conduction band and the Bi-related edge states, which are not temperature sensitive. That is why there is no obvious red shift found in the response curve as temperature increases. As the temperature further increases beyond 220K, the valence band of InGaAsBi further upshifts and eventually raise above the Bi-related edge states, then the extension of cutoff wavelength becomes perceptible. It is also noted that the responsivity decreased down to about 220K and then start to increase again up to 300K as show in the inset of Fig.4. As mentioned above, when the temperature is below 220K, Bi-related edge state is above valence band of InGaAsBi, the absorption happens as the electrons in the Bi-related edge state transited up to conduction band. As the temperature increases, the defects become to be more effective recombination centers, and the carrier collection efficiency drops which results in a decreased responsivity. When the temperature is above 220K, valence band is uplifted above Bi-related edge state, as the temperature increases, the shrink of InGaAsBi bandgap causes an extension of the cutoff wavelength and a larger absorption coefficient, which enhances the responsivity of the device.

PHOTOLUMINESCENCE CHARACTERIZATION

In order to probe the optical property of the InGaAsBi materials and verify the explanation in the last section, the temperature varied photoluminescence measurement on the wafer with p-type cap layer removed was carried out. Here, 532nm laser was used as an excitation source, and emission spectra were detected by In-rich InGaAs detector equipped in Fourier transform infrared spectrometer (FTIR). Fig. 5(a) shows the temperature dependent PL spectra from 20K to160K. Two emission peaks were observed one peaked around 2.25μm, (marked as peak A), another one peaked around 2.45μm (marked as peak B). The positions of both two peaks are temperature-insensitive, which are due to the energy transition between Bi-related deep localizations in bandgap and conduction band edge (CBE) [26, 27]. As the temperature increases, compare with the upward movement of VBE, the downshift of CBE is much smaller and the Bi-related bound states are expected to be fixed above the valence band edge (VBE) as well. That is why no obvious shift was observed in the PL spectra as temperature increases. Moreover, this peak A at wavelength of 2.25μm explains the cutoff wavelength behaviors of the InGaAsBi PD as temperature changes from 77K to 300K. To further probe the emission with longer wavelength, modulated infrared PL technique based on a step-scan FTIR spectrometer [28, 29] with a liquid nitrogen cooled InSb detector was used to retest the PL spectra, as shown in Fig. 5(b). Here two peaks at 2.6μm (0.48 eV) and 3.54μm (0.35 eV) were found and marked as peak C and peak D. Compare with peak C, the intensity of peak D is weaker and only can be distinguished at low temperature. It is also noted that both peaks are temperature-insensitive, which indicate two other Bi-related deep levels within the bandgap. Here, we used two different photodetectors for photoluminescence measurement in order to have the best signal to noise ratio in the whole spectrum. In Fig. 5(a), the PL signals cut off at 2.6μm, which is limited by the cutoff wavelength of 2.6μm for InGaAs photodetector. Moreover, it is also noted that the peak A (around 2.25μm) observed by InGaAs photodetector in Fig.4(a) is not visible in spectrum detected by InSb photodetector as shown in Fig. 5(b). It is believed to be due to the weak response of InSb photodetector in that wavelength range.

LOW FREQUENCY NOISE SPECTROSCOPY

Noise in a photodetector is an import figure of merit, which often limit the performance of photodetectors. White noise and low frequency noise (LFS) are the main noise in photodetectors. The current power spectral density (PSD) is often observed frequency dependent in low frequency region (below several MHz), which is called low frequency noise. In high frequency region, the PSD becomes frequency-independent white noise, generally limited by thermal and shot noise. LFN includes 1/f noise and generation -recombination (G-R) noise, where the PSD of 1/f noise is inversely proportional to frequency and the G-R noise is induced by the random capture and emission carrier from deep levels in the bandgap. The G-R noise arising from deep levels appeared as Lorentzian peaks superimposed on the noise spectra. The detail of LFN measurement can refer to the reference [30-33].

The current noise spectral density is mathematically expressed as [33]:

$$S_I = \sum \frac{A_i \tau_i}{1+(2\pi f \tau_i)^2} + \frac{B}{f} + C \quad (1)$$

Where $A_i$ and B are the amplitudes of the G-R and 1/f process, $\tau_i$ is the time constant of the G-R center, C is the white noise related constant. By plotting the $f*S_I$ versus f, the Lorentzian peaks would be more visible, and the time constant $\tau_i$ under different temperatures can be extracted by Lorentzian fitting. In addition, time constant $\tau_i$ can be written as $\tau_i=1/(2\pi f_c)$, $f_c$ is corner frequency where Lorentzian appear as symmetric peak . In a n-type material, we have time constant [34]:

$$\tau = (v_{th} \sigma N_c)^{-1} \exp((E_C - E_T)/k_B T) \quad (2)$$

Where $v_{th}$ is electron's thermal velocity, $\sigma$ is capture cross section, $E_T$ is deep level position, $E_c$ and $N_c$ are the conduction band edge and effective density of states, $v_{th}$ and $N_c$ are $T^{1/2}$, $T^{3/2}$ dependent, respectively. From the Arrhenius plot of $\ln(\tau T^2)$ versus 1000/T, the energy level and capture cross section of the deep levels can be obtained based on the slope and intercept.

To further investigate the defect information in InGaAsBi, low frequency noise spectroscopy was applied to the InGaAsBi photodiodes, Device with diameter of 300 μm was measured under the temperature from 77K to 300K with an increasing step of 6K, at the bias voltage of 100 mV. Fig. 6 shows the measured noise spectra under different temperature ranges, although the noise spectra under some temperatures look a little noisy, the Lorentzian peak is observed evidently. The time constant corresponding to the carrier lifetime at each temperature was extracted by Lorentzian fitting. According to the Arrhenius plot in Fig. 7, three deep levels with activation energy of 0.33 eV ($E_1$), 0.14 eV ($E_2$) and 0.51 eV ($E_3$) were found, the associated capture cross section of these three deep levels are $3.5\times10^{-11}$ cm$^2$, $6.3\times10^{-20}$ cm$^2$ and $8.7\times10^{-11}$cm$^2$, respectively. As can be seen in Fig. 6, the Lorentzian peaks of the first observed deep level maintain from 137K to 179K, the time constants extracted from these eight temperatures are expected to be fitted linearly. It is noted that the last three points deviated from the linear region, which was excluded for linear fitting as shown in Fig. 7. It is unclear why these points deviated from linear region, and some future work may be necessary to investigate the reason behind.

As mentioned earlier, $A_i$ in Eq. (1) is the amplitude of the G-R process and is proportional to the factor F(1-F). Here F is the equilibrium deep level occupation and $F = [1+\exp{(E_T-E_F)/kT}]^{-1}$ [35], where $E_T$ and $E_F$ are the deep level and Fermi level positions. One can see that when the deep level located near the Fermi energy, the G-R noise will be maximized in the noise spectra. Under a certain bias, the G-R noise could only be observed at a temperature which shifts the quasi-Fermi level close to the deep level. Generally speaking, as the temperature changes from low to high, the fermi level would shift monotonically. Therefore, it can be inferred that the energy different between deep level and conduction band should also change monotonically. Thus, the three deep levels found above should be at $E_c$ -0.33 eV, $E_v$ +0.14 eV, and $E_c$ -0.51 eV. Recalling the emission energy (0.55 eV, 0.48 eV, 0.35 eV) found in PL spectra, two deep levels (0.35 eV, 0.48 eV) derived from the PL spectra are close to the energy levels found by LFNS which are located at $E_c$ -0.33 eV and $E_c$ -0.51 eV [36]. Ideally, the edge state level with activation energy of 0.55eV found in PL spectra could also been found in LFNS measurement, however, as temperature increases, the valence band is shifted above the edge state level, which make the it challenging to be probed. Moreover, it is also noted that emission between valence band and the defect level with activation energy of 0.14eV found LFNS, corresponding to wavelength of around 8.8μm, is not observed in PL spectra. This may be due to the fact that the cutoff wavelength of the InSb detector used in PL system is only around 5.5μm. Along with the data we got, the energy band schematic of the InGaAaBi material with the trap information can be roughly deduced in Fig. 8.

## CONCLUSIONS

In summary, a comprehensive study was carried out to investigate the deep levels in the InP based InGaAsBi photodiodes. Three deep levels located at $E_c$ -0.33 eV, $E_v$ +0.14 eV, and $E_c$ -0.51 eV were found by LFNS. Two of them correlate well with the PL spectra of InGaAsBi, where three temperature-insensitive peaks at energy of 0.35 eV, 0.48 eV, 0.55 eV was observed. The additional peak around 0.55eV in PL spectra is consistent to the abnormal redshift behavior of optical response. These defects information found in this work is significant and provides a method for the characterization of deep levels in InGaAsBi, which may help further optimize and improve the InGaAsBi material quality.

## ACKNOWLEDGEMENTS

This study is supported by Shanghai Sailing Program (17YF1429300), ShanghaiTech University startup funding (F-0203-16-002) and the National Natural Science Foundation of China (61775228).

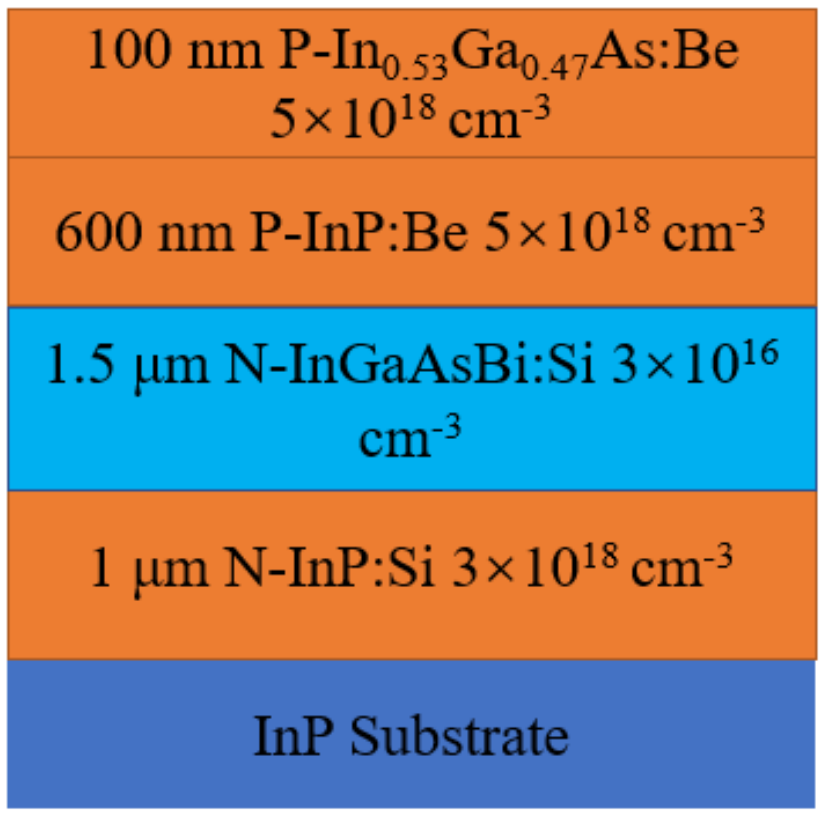

FIG. 1. Schematic of the p-i-n InGaAsBi photodetector.

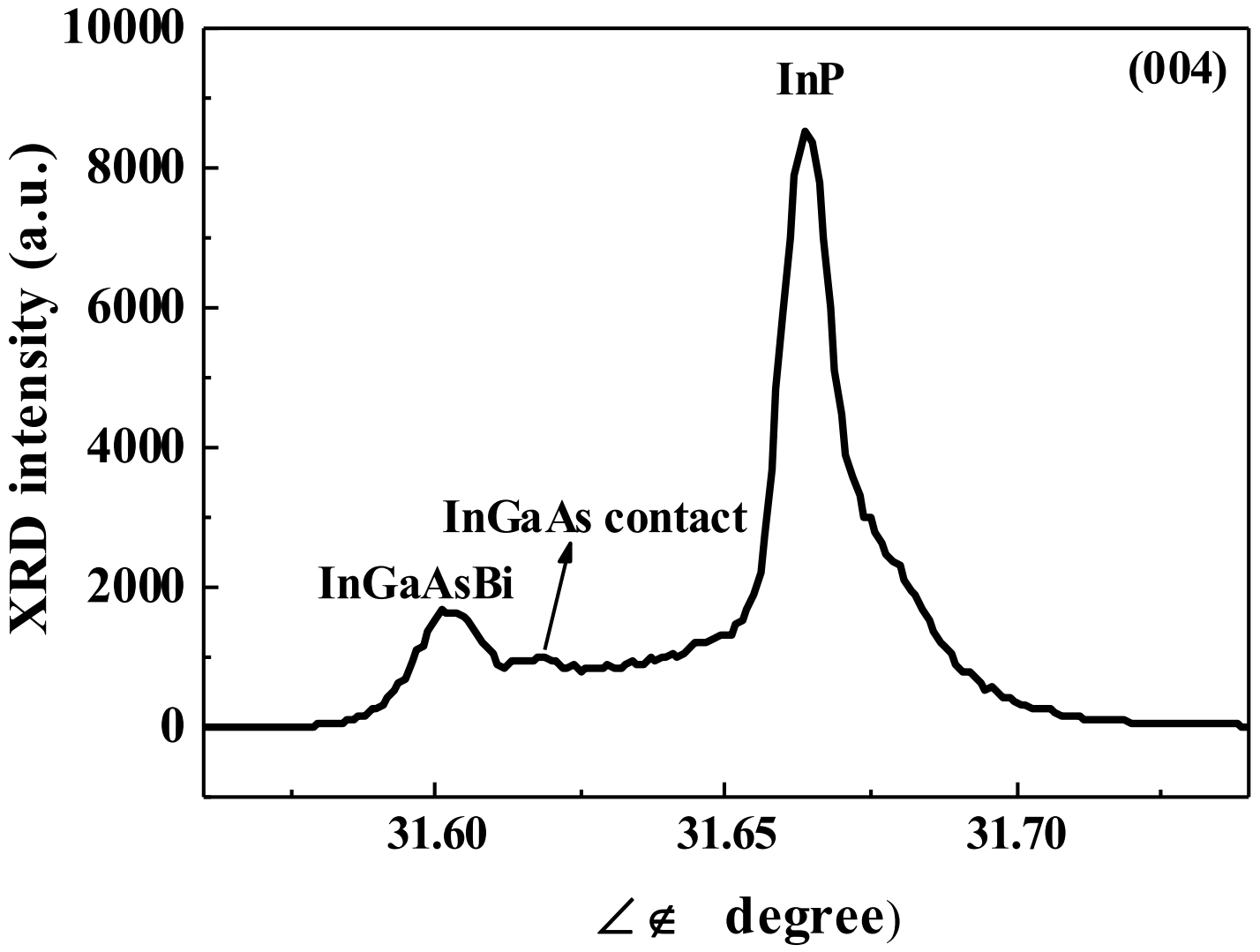

FIG.2 HRXRD (004) ω/2θ scan curves of the detector sample.

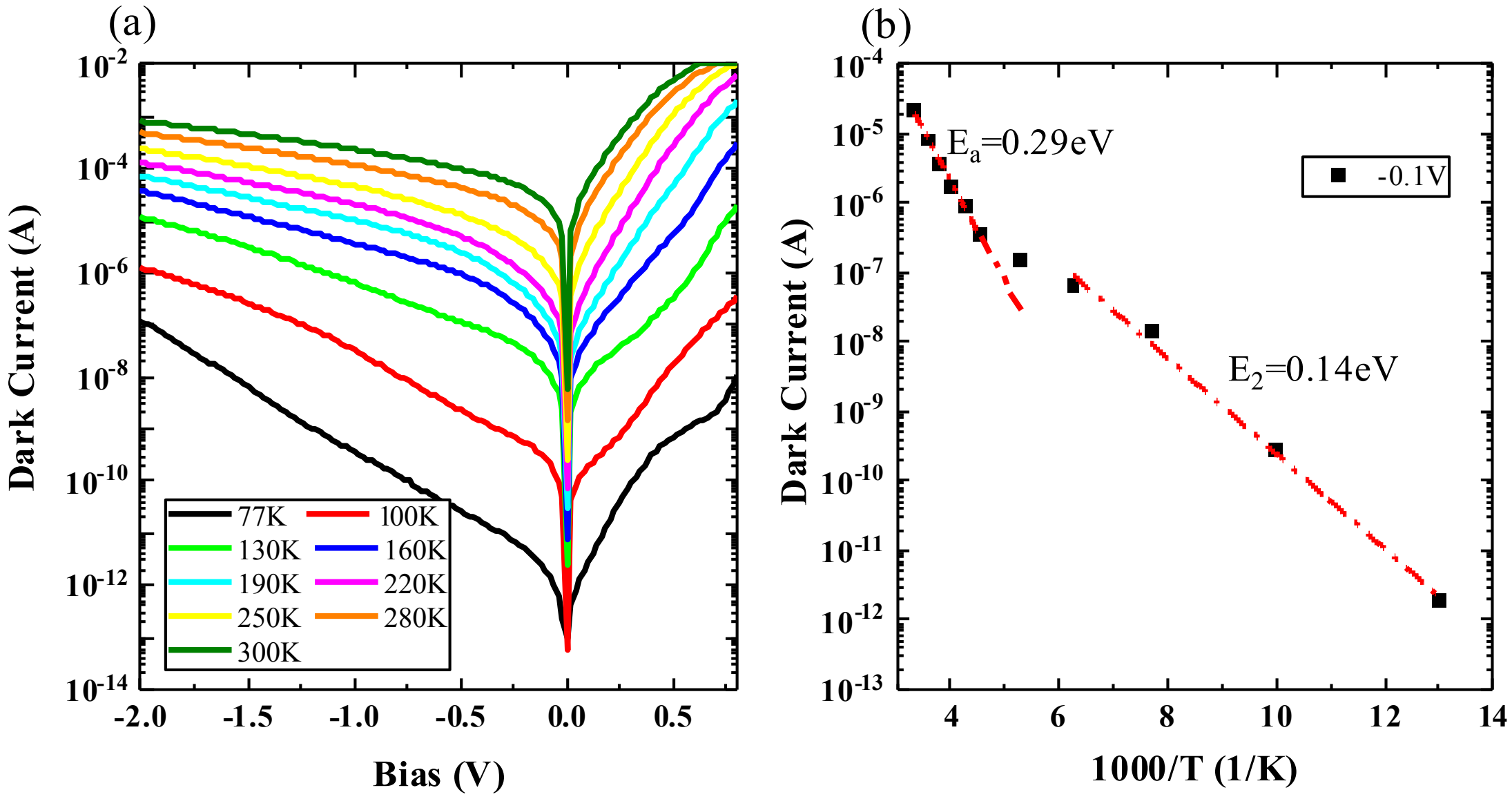

FIG. 3. (a) Dark I-V curves as a function of temperature of a 300μm diameter photodetector. (b) Arrhenius plot of the temperature dependent dark current at -0.1V.

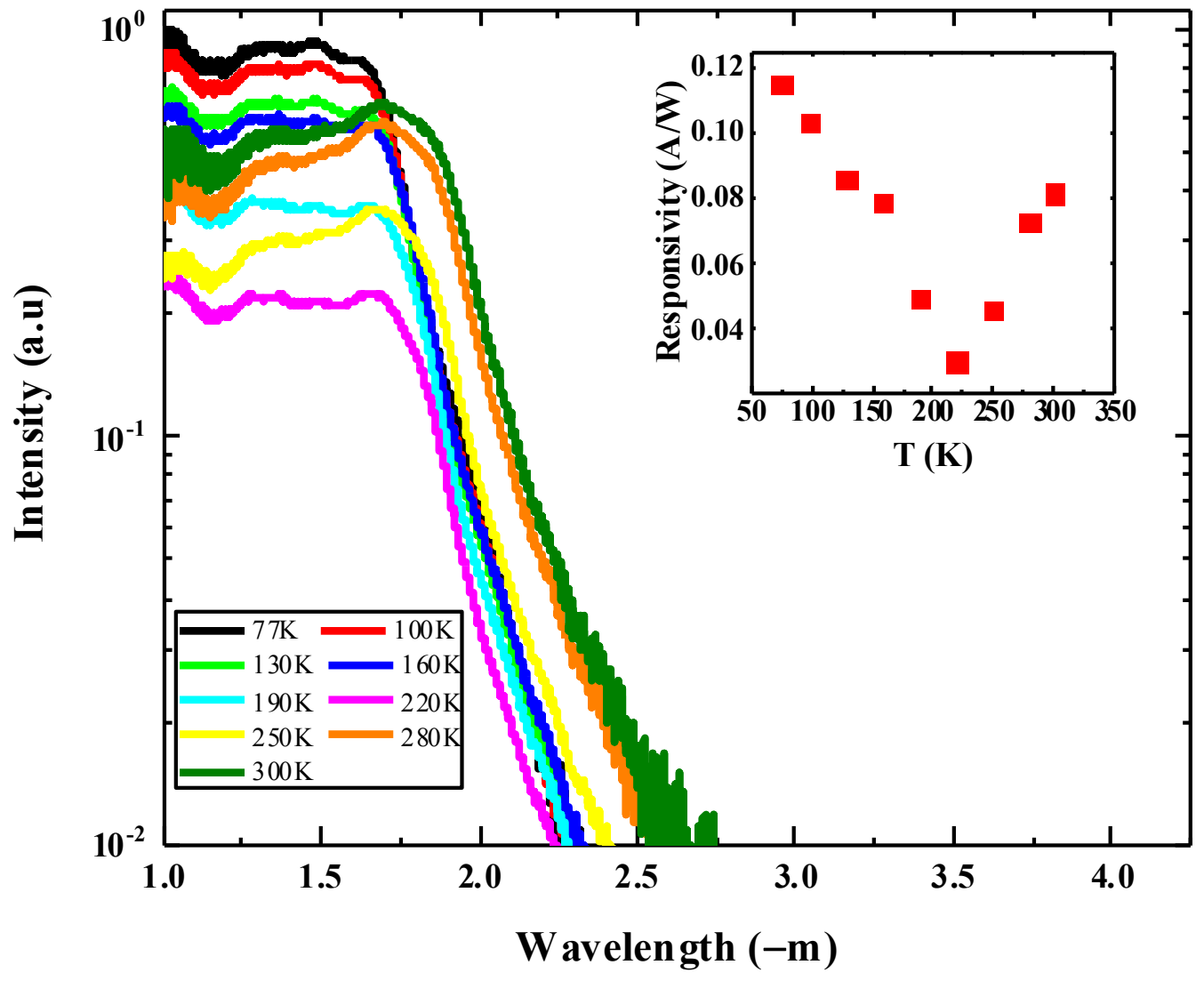

FIG. 4. Normalized responsivity of InGaAsBi photodetector from 77K to 300K under zero bias. Insert: responsivity of the device at 1.6μm at different temperatures under zero bias.

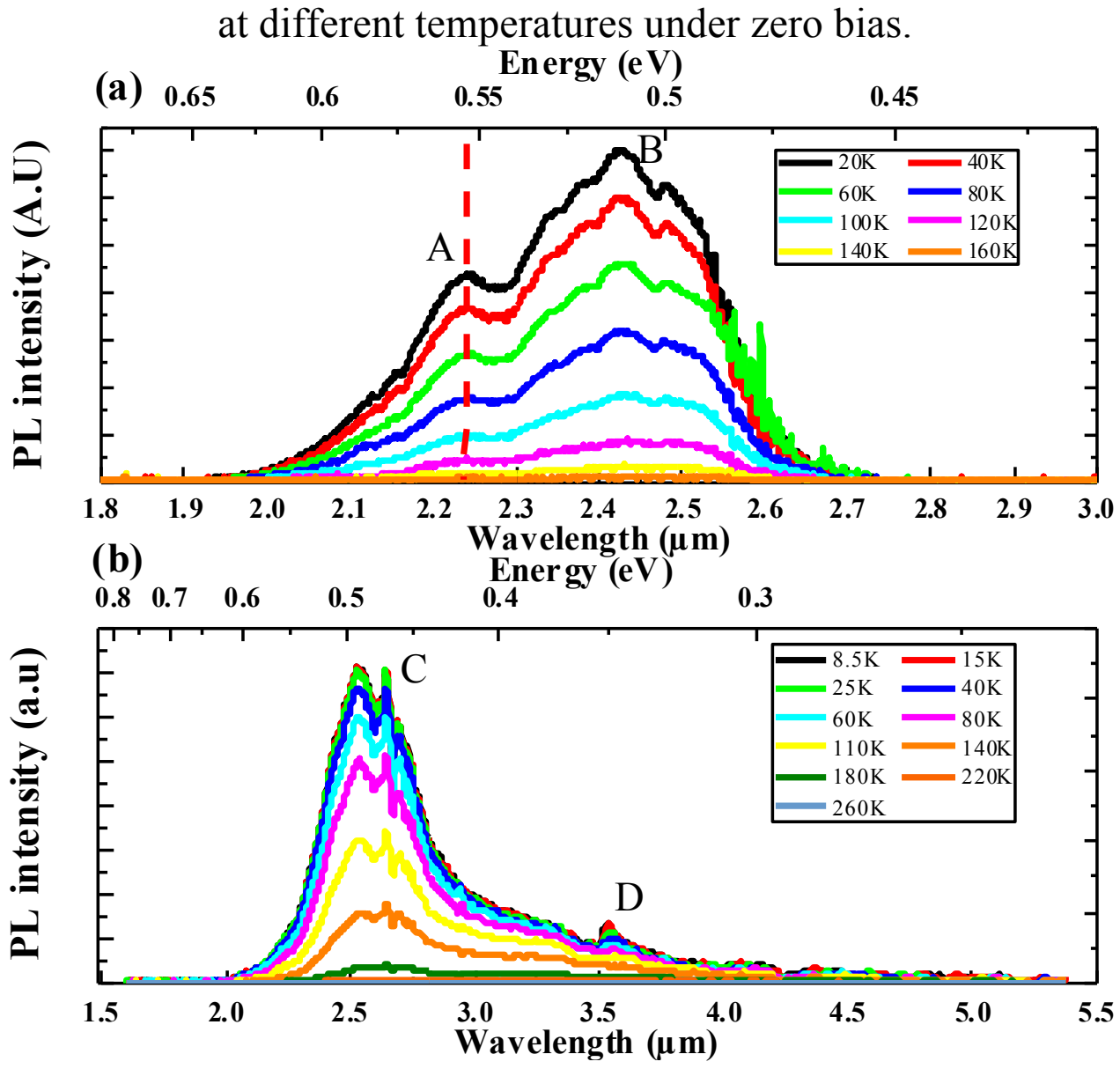

FIG. 5. Temperature dependent PL spectra of the InGaAsBi layer (a) detected by InGaAs detector in FTIR spectrometer and (b) detected by InSb detector in FTIR spectrometer.

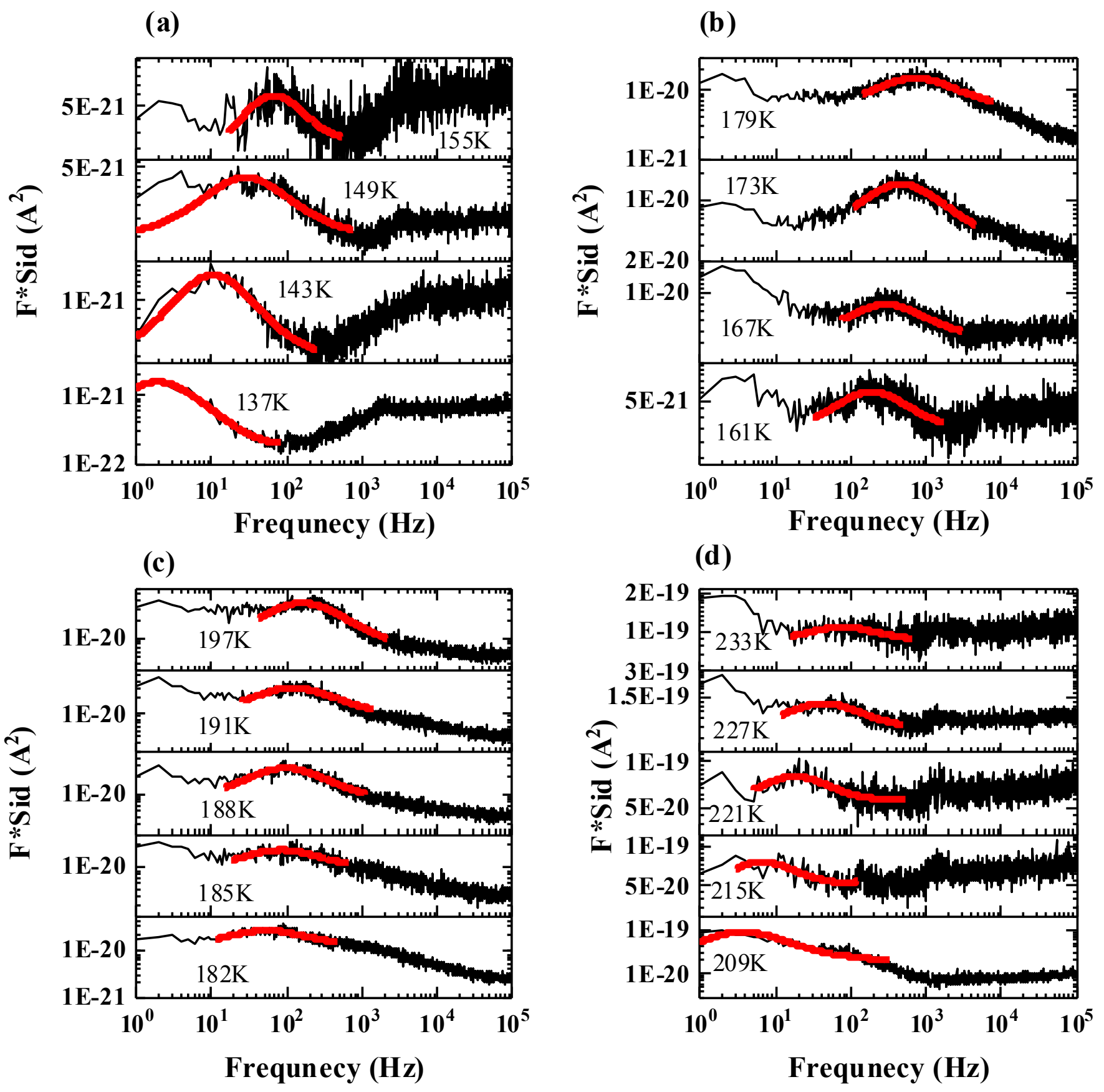

FIG.6. The measured noise spectra under different temperatures with Lorentzian fitting.

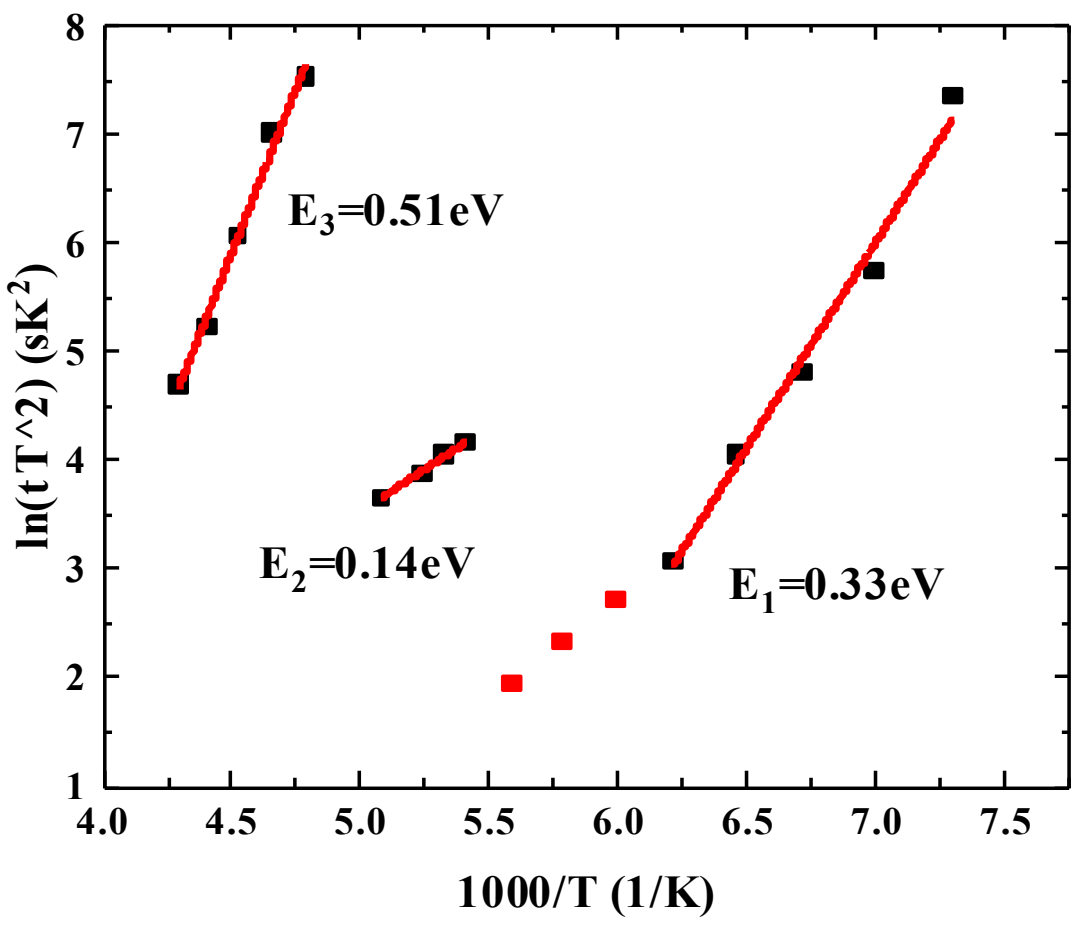

FIG. 7. Arrhenius plot of the deep levels in InGaAsBi PD.

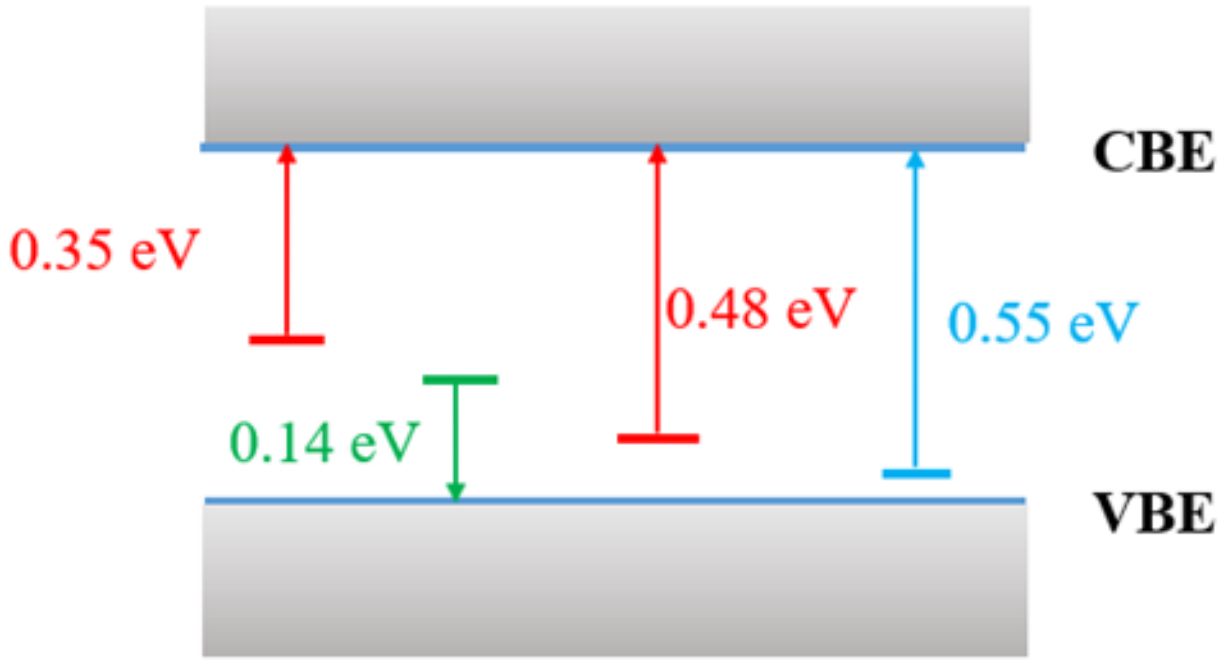

FIG. 8. Energy band schematic of the InGaAsBi PD. The blue solid lines represent for CBE and VBE, the gray shade areas represent for the conduction band and valence band; The short lines with different colors in bandgap represent for discrete energy levels, the red short lines depict the deep levels found by both LFNS and PL spectra, the green short line depicts the deep level found by LNFS only, and the blue short line depicts the edge state found by PL spectral only; all the corresponding energy of these discrete energy levels are listed and the value of energy mainly refer to PL results. (The positions of the energy levels are not drawn to scale)

**REFERENCES**

1. T.-P. Lee, C. Burrus, and A. Dentai, "InGaAs/InP pin photodiodes for lightwave communications at the 0.95-1.65 µm wavelength," *IEEE J. Quantum Electron.* 17(2), 232-238, (1981).
2. Y. Arslan, F. Oguz, and C. Besikci, "Extended wavelength SWIR InGaAs focal plane array: characteristics and limitations," *Infrared Phys. Technol* 70134-137, (2015).
3. X. Li, H. Gong, J. Fang, H. Tang, S. Huang, T. Li, and Z. Huang, "The development of InGaAs short wavelength infrared focal plane arrays with high performance," *Infrared Phys. Technol* 80112-119, (2017).
4. P. Jurczak, K. A. Sablon, M. Gutiérrez, H. Liu, and J. Wu, "2.5-µm InGaAs photodiodes grown on GaAs substrates by interfacial misfit array technique," *Infrared Phys. Technol* 81320-324, (2017).
5. N. Cohen and O. Aphek, "Extended wavelength SWIR detectors with reduced dark current," in *Infrared Technology and Applications XLI*, 2015, vol. 9451, p. 945106: International Society for Optics and Photonics.
6. Y. Uliel, D. Cohen-Elias, N. Sicron, I. Grimberg, N. Snapi, Y. Paltiel, and M. Katz, "InGaAs/GaAsSb Type-II superlattice based photodiodes for short wave infrared detection," *Infrared Phys. Technol* 8463-71, (2017).
7. R. Sidhu, N. Duan, J. C. Campbell, and A. L. Holmes, "A long-wavelength photodiode on InP using lattice-matched GaInAs-GaAsSb type-II quantum wells," *IEEE Photonics Technol. Lett.* 17(12), 2715-2717, (2005).
8. H. Inada, K. Miura, Y. Nagai, M. Tsubokura, A. Moto, Y. Iguchi, and Y. Kawamura, "Low dark current SWIR photodiode with InGaAs/GaAsSb type II quantum wells grown on InP substrate," in *2009 IEEE International Conference on Indium Phosphide & Related Materials*, 2009, pp. 149-152: IEEE.
9. T. Kawahara, K. Machinaga, B. Sundararajan, K. Miura, M. Migita, H. Obi, T. Fuyuki, K. Fujii, T. Ishizuka, and H. Inada, "InGaAs/GaAsSb type-II quantum-well focal plane array with cutoff-wavelength of 2.5 µm," in *Quantum Sensing and Nano Electronics and Photonics XIV*, 2017, vol. 10111, p. 1011115: International Society for Optics and Photonics.

10. C. Jin, F. Wang, Q. Xu, C. Yu, J. Chen, and L. He, "Beryllium compensation doped InGaAs/GaAsSb superlattice photodiodes," *J. Cryst. Growth* 477100-103, (2017).
11. B. Chen, W. Y. Jiang, J. Yuan, A. L. Holmes, and B. M. Onat, "Demonstration of a Room-Temperature InP-Based Photodetector Operating Beyond 3 μm," *IEEE Photonics Technol. Lett.* 23(4), 218-220, (2011).
12. B. Chen, W. Jiang, and A. Holmes, "Design of strain compensated InGaAs/GaAsSb type-II quantum well structures for mid-infrared photodiodes," *Opt. Quantum Electron.* 44(3-5), 103-109, (2012).
13. B. Chen, W. Jiang, J. Yuan, A. L. Holmes, and B. M. Onat, "SWIR/MWIR InP-based pin photodiodes with InGaAs/GaAsSb type-II quantum wells," *IEEE J. Quantum Electron.* 47(9), 1244-1250, (2011).
14. . *http://www.hamamatsu.com.cn/product/21749.html*.
15. L. Yue, Y. Song, X. Chen, Q. Chen, W. Pan, X. Wu, J. Liu, L. Zhang, J. Shao, and S. Wang, "Novel type II InGaAs/GaAsBi quantum well for longer wavelength emission," *Journal of Alloys and Compounds* 695753-759, (2017).
16. B. Chen, "Active Region Design and Gain Characteristics of InP-Based Dilute Bismide Type-II Quantum Wells for Mid-IR Lasers," *IEEE Trans. Electron Devices* 64(4), 1606-1611, (2017).
17. C. A. Broderick, S. Jin, I. P. Marko, K. Hild, P. Ludewig, Z. L. Bushell, W. Stolz, J. M. Rorison, E. P. O'Reilly, and K. Volz, "GaAs 1− x Bi x/GaN y As 1− y type-II quantum wells: novel strain-balanced heterostructures for GaAs-based near-and mid-infrared photonics," *Sci. Rep.* 746371, (2017).
18. B. Chen, "Optical gain analysis of GaAs-based InGaAs/GaAsSbBi type-II quantum wells lasers," *Opt Express* 25(21), 25183-25192, (2017).
19. M. K. Rajpalke, W. Linhart, M. Birkett, K. Yu, J. Alaria, J. Kopaczek, R. Kudrawiec, T. Jones, M. Ashwin, and T. Veal, "High Bi content GaSbBi alloys," *J. Appl. Phys.* 116(4), 043511, (2014).
20. K. Alberi, O. Dubon, W. Walukiewicz, K. Yu, K. Bertulis, and A. Krotkus, "Valence band anticrossing in Ga Bi x As 1− x," *Appl. Phys. Lett.* 91(5), 051909, (2007).
21. R. Butkutė, V. Pačebutas, B. Čechavičius, R. Nedzinskas, A. Selskis, A. Arlauskas, and A. Krotkus, "Photoluminescence at up to 2.4 μm wavelengths from GaInAsBi/AlInAs quantum wells," *J. Cryst. Growth* 391116-120, (2014).
22. S. Tixier, M. Adamcyk, T. Tiedje, S. Francoeur, A. Mascarenhas, P. Wei, and F. Schiettekatte, "Molecular beam epitaxy growth of GaAs 1− x Bi x," *Appl. Phys. Lett.* 82(14), 2245-2247, (2003).
23. Y. Gu, Y. Zhang, X. Chen, Y. Ma, S. Xi, B. Du, and H. Li, "Nearly lattice-matched short-wave infrared InGaAsBi detectors on InP," *Appl. Phys. Lett.* 108(3), 032102, (2016).
24. S. Francoeur, S. Tixier, E. Young, T. Tiedje, and A. Mascarenhas, "Bi isoelectronic impurities in GaAs," *Phys. Rev. B: Condens. Matter* 77(8), 085209, (2008).
25. R. Kudrawiec, M. Syperek, P. Poloczek, J. Misiewicz, R. Mari, M. Shafi, M. Henini, Y. G. Gobato, S. Novikov, and J. Ibanez, "Carrier localization in GaBiAs probed by photomodulated transmittance and photoluminescence," *J. Appl. Phys.* 106(2), 023518, (2009).
26. X. Chen, Y. Gu, Y. Zhang, S. Xi, B. Du, Y. Ma, W. Ji, and Y. Shi, "Characteristics of InGaAsBi with various lattice mismatches on InP substrate," *AIP Advances* 6(7), 075215, (2016).
27. K. Alberi, B. Fluegel, D. A. Beaton, M. Steger, S. A. Crooker, and A. Mascarenhas, "Origin of deep localization in GaA s 1− x B i x and its consequences for alloy properties," *Physical Review Materials* 2(11), 114603, (2018).

28. J. Shao, W. Lu, X. Lü, F. Yue, Z. Li, S. Guo, and J. Chu, "Modulated photoluminescence spectroscopy with a step-scan Fourier transform infrared spectrometer," *Rev. Sci. Instrum.* 77(6), 063104, (2006).
29. Z. Deng, B. Chen, X. Chen, J. Shao, Q. Gong, H. Liu, and J. Wu, "Optical properties of beryllium-doped GaSb epilayers grown on GaAs substrate," *Infrared Phys. Technol* 90115-121, (2018).
30. W. Chen, B. Chen, J. Yuan, A. Holmes, and P. Fay, "Bulk and interfacial deep levels observed in In0.53Ga0.47As/GaAs0.5Sb0.5 multiple quantum well photodiode," *Appl. Phys. Lett.* 101(5), (2012).
31. W. Chen, B. Chen, A. Holmes, and P. Fay, "Investigation of traps in strained-well InGaAs/GaAsSb quantum well photodiodes," *Electron. Lett.* 51(18), 1439-1440, (2015).
32. J. Huang, Y. Wan, D. Jung, J. Norman, C. Shang, Q. Li, K. M. Lau, A. C. Gossard, J. E. Bowers, and B. Chen, "Defect characterization of InAs/InGaAs quantum dot pin photodetector grown on GaAs-on-V-grooved-Si substrate," *ACS Photonics* (2019).
33. B. K. Jones, "Low-frequency noise spectroscopy," *IEEE Trans. Electron Devices* 41(11), 2188-2197, (1994).
34. L. Ciura, A. Kolek, A. Kębłowski, D. Stanaszek, A. Piotrowski, W. Gawron, and J. Piotrowski, "Investigation of trap levels in HgCdTe IR detectors through low frequency noise spectroscopy," *Semicond. Sci. Technol.* 31(3), 035004, (2016).
35. M. Levinshtein and S. Rumyantsev, "Noise spectroscopy of local levels in semiconductors," *Semicond. Sci. Technol.* 9(6), 1183, (1994).
36. J. Wróbel, Ł. Ciura, M. Motyka, F. Szmulowicz, A. Kolek, A. Kowalewski, P. Moszczyński, M. Dyksik, P. Madejczyk, and S. Krishna, "Investigation of a near mid-gap trap energy level in mid-wavelength infrared InAs/GaSb type-II superlattices," *Semicond. Sci. Technol.* 30(11), 115004, (2015).